\documentstyle[12pt]{article}

\begin{document}

\newcommand{\beq}{\begin{eqnarray}}
\newcommand{\eeq}{\end{eqnarray}}
\newcommand{\nn}{\noindent}
\newcommand{\non}{\nonumber}
\newcommand{\ee}{e^+ e^-}
\newcommand{\ov}{\overline}
\newcommand{\st}{\stackrel{-}}
\newcommand{\cpb}{\overline{\cal P}}
\newcommand{\cp}{{\cal P}}
\newcommand{\lra}{\longrightarrow}
\newcommand{\ra}{\rightarrow}
\newcommand{\La}{{\rm log}\frac{1-\mu^2+v_T}{v_T}}
\newcommand{\SM}{{\rm Standard \ Model}}
\newcommand{\SMG}{ {\rm	SU(3)_C	\times SU(2)_L \times U(1)_Y }}
\newcommand{\sw}{s_W^2}
\newcommand{\sbb}{\sin \beta}
\newcommand{\cb}{\cos \beta}
\newcommand{\s}{\\ \vspace*{-3mm} }

\renewcommand{\thefootnote}{\fnsymbol{footnote} }
\pagestyle{empty}

\nn \hspace*{12cm} UdeM-LPN-TH-93-157 \\
\hspace*{12cm} July 1993 \\

\vspace*{1.8cm}

\centerline{\large{\bf New Fermions at e$^+$e$^-$ Colliders:}}

\vspace*{0.5cm}

\centerline{\large{\bf I.~Production and Decay. }}

\vspace*{1.5cm}

\centerline{\sc A.~Djouadi\footnote{NSERC Fellow.}}

\vspace*{1cm}

\centerline{Laboratoire de Physique Nucl\'eaire, Universit\'e de Montr\'eal,
 Case 6128 Suc.~A,}
\centerline{H3C 3J7 Montr\'eal PQ, Canada.}

\vspace*{2cm}

\begin{center}
\parbox{14cm}
{\begin{center} ABSTRACT \end{center}
\vspace*{0.2cm}

\nn We analyze the production in e$^+$e$^-$ collisions of new heavy fermions
stemming from extensions of the Standard Model. We write down the most general
expression for the production of two heavy fermions and their subsequent
decays,
allowing for the polarization of the e$^+$e$^-$ initial state and taking into
account the final polarization of the fermions. We then discuss the various
decay modes including cascade and three body decays, and the production
mechanisms, both pair production and single production in association with
ordinary fermions.}

\end{center}

\renewcommand{\thefootnote}{\arabic{footnote} }
\setcounter{footnote}{0}

\newpage
\pagestyle{plain}

\subsection*{1.~Introduction}

\renewcommand{\theequation}{1.\arabic{equation}}
\setcounter{equation}{0}

Despite its tremendous sucess in describing all experimental data available
today, the Standard Model of the electromagnetic, weak and strong interactions
based on the gauge symmetry $\SMG$, is widely believed not to be the ultimate
truth. Besides the fact that it has too many parameters which are merely
incorporated by hand, the $\SM$ does not unify the electroweak  and strong
forces in a satisfactory way since the coupling constants of these interactions
are different and seem to be independent. Therefore, one expects the existence
of a more fundamental theory which describes the three forces within the
context of a single gauge group and hence, with only one coupling constant.
This grand unified theory will be based on a gauge group containing $\SMG$ as a
subgroup and will reduce to this symmetry at present energies. \s

\nn The grand unified groups [1--4] provide fermion representations in which a
complete generation of Standard Model quarks and leptons can be naturally
embedded.  In most of the cases these representations are large enough to
accomodate additional new fermions which, in fact, are needed to have
anomaly--free theories.  It is conceivable that these new fermions, if for
instance they are protected by some symmetry, acquire masses not much larger
than the Fermi scale. This is	very likely, and even necessary if the new
gauge  bosons which are generic predictions of the unified theories, are
relatively light \cite{R5}. \s

\nn Besides the SU(5) group \cite{R1} [the simplest Lie group containing $\SMG$
as a subgroup and with two representations to accomodate the 15 Standard Model
fermions] which has no room for ``light" new fermions or gauge bosons, the
SO(10) group \cite{R3} has received much attention. It is the simplest group in
which the 15 Weyl spinors of each $\SM$ generation of fermions can be embedded
into a single multiplet.  This representation has dimension {\bf 16} and, to
have an anomaly--free theory, contains a right--handed Majorana neutrino. In
fact, heavy isosinglet neutrinos have been discussed in various models, such as
left--right symmetric models \cite{R6}, in attempts to explain the small masses
of the three observed neutrinos. \s

\nn Another popular unifying group is E$_6$ \cite{R4} which contains SU(5) and
SO(10) as subgroups and is the next anomaly--free choice after SO(10). The
interest in E$_6$ is mainly due the fact that superstring theories, which
attempt to unify all fundamental forces including gravity, suggest that this
symmetry is an acceptable four	dimensional field theoretical limit \cite{R7}.
In E$_6$, each quark--lepton generation lies in the representation of dimension
{\bf 27}; to complete this representation, twelve new fields are needed in
addition the $\SM$ fermion fields. For each family	one has	two additional
isodoublets of leptons, two isosinglets neutrinos [which can be either of the
Dirac or Majorana type] and an isosinglet quark with	charge $-1/3$. \s

\nn Several other gauge	groups have been considered with various theoretical
motivations and most of them predict the existence of new fermions. For
instance, schemes of grand unification based on large orthogonal groups have
been proposed to explain the origin of parity violation in weak interactions
\cite{R8}: in these models, weak interactions are parity symmetric but fermions
with left--handed and right--handed couplings aquire different masses. They
predict a rich spectrum of fermions, called mirror fermions \cite{R9}, which
have the opposite	chiral properties of the ordinary ones. In the simplest
version of these models	\cite{R10}, the	gauge symmetry and the symmetry
breaking pattern are the same as in the	Standard Model:	one simply adds
to	the spectrum of	the latter three families of heavy fermions with
opposite chiralities. Theoretical arguments based on the unitarity of
scattering amplitudes \cite{R11} suggest that the masses of these mirror
fermions should not exceed a few	hundred	GeV. \\

\nn The	direct search for these	new fermions and, in case of discovery, the
study of their basic properties, will be a major goal of the next generation of
accelerators. In this paper, we analyze in detail the production of these new
fermions at $\ee$ colliders. \\

\nn The new leptons and quarks will mix with the ordinary fermions of the
Standard Model \cite{R12,R13}. This mixing will give rise to new currents which
determine to a large extent their decay properties and allows for new
production mechanisms. If the new particles have non--zero electromagnetic and
weak charges, they can be pair produced if their masses are smaller than the
beam energy. In general the reactions are built--up by a superposition of
photon and $Z$ boson exchange [additional contributions could come from extra
gauge bosons if their masses are not much larger than the c.m.~energy of the
collider]. The cross sections are large \cite{R14} and, up to phase space
suppression factors, of the order of the point-like QED cross section for muon
pair production. \s

\nn Fermion mixing allows an additional production mechanism for the new
fermions: single production in association with their light partners. In the
case of quarks [and for second and third generation new leptons if
inter--generational mixing is neglected] the production process is mediated by
$s$--channel gauge boson exchange; since the mixing angles are restricted to
values smaller than ${\cal O} (10^{-1})$ by present experimental data
\cite{R13}, the cross sections are rather small. But in the case of [the first
generation] heavy leptons, additional $t$--channel exchanges, $W$ exchange for
neutral leptons and $Z$ exchange for charged leptons, are present increasing
the cross sections by several orders of magnitude. This results in large
production rates \cite{R14} which permit to probe lepton masses close to the
total c.m. energy. \s

\nn The new fermions will decay through mixing into light fermions and gauge
bosons. Depending on the particle masses, the gauge bosons can be real or
virtual and will decay into light quarks and leptons. Therefore, in the
production of the new fermions the final states are rather complicated: six
particles in the case of pair production and four particles for single
production. However, it is very important to have at hand the information on
the correlation between all the particles involved in the process. Indeed,
these correlations will be very helpful to optimize the experimental cuts which
permit to suppress the various backgrounds without affecting drastically the
signal cross sections. Furthermore, they permit to discriminate between
particles of different nature [e.g.~Majorana or Dirac neutrinos] or with
different couplings [vector, ``mirror" or standard couplings in pair
production, or with left and right--handed mixing in single production] and
therefore shed some light on the origin of the new fermion. \\

\nn Several analyses of the production of new fermions in $\ee$ collisions have
been conducted in the recent years [14-20] in various special cases.
In this paper we will extend on these analyses in the following ways: \s

\nn $(i)$ We present the most general expressions for the production of two
heavy fermions, with different masses to allow for the production of two
different fermions and to treat the cases of single and pair
production in the same footing, of any flavor and for arbitrary couplings of
the new fermions, including all possible channels and the polarization of the
initial $\ee$ state; complete formulae are given for angular distributions and
total cross sections. \s

\nn $(ii)$ We give the most general expression of the decay of the heavy
fermions into three body final states, allowing for cascade decays, including
all channels and the possibility of off-shell intermediate vector bosons;
complete formulae for the angular and energy distributions of the final decay
products as well as the total decay widths of the heavy fermions are given. \s

\nn $(iii)$ We systematically take into account the polarization of the heavy
fermions in both the production and the decay processes; due to the
factorization of the two sequences, the information on the final polarization
permits an easy reconstruction of the full correlations between the initial
state and the final particles from the decays of the heavy fermions. \\

\nn Note that since our analytical results are general, they can also be used
to discuss, at $\ee$ colliders, the details of top quark pair production, or
the production of heavy fermions of a fourth generation with a heavy
neutrino.  \\

\nn The paper is organized as follows. In the next section, we summarize the
interactions of the new fermions. In section 3, we analyze the general case of
the production of two heavy fermions and discuss the cases of pair production
and the single production in association with ordinary fermions. In section 4,
we discuss in detail the decay modes of the new fermions including the cascade
decays and the three body decays with off--shell gauge bosons. Section 5
contains our conclusions. For completeness, we summarize in the Appendix the
formalism for combining the spin--dependent cross sections and decay
distributions which permits to obtain the full correlations between all the
particles involved in the process.

\subsection*{2.~Interactions}

\renewcommand{\theequation}{2.\arabic{equation}}
\setcounter{equation}{0}

The new fermions couple to the photon and the electroweak gauge bosons $W/Z$
with full strength, except for singlet neutrinos which have zero
electromagnetic
and weak charges and therefore couple to the latter only through mixing as will
be discussed later. They similarly couple to extra gauge bosons
when the latter are present. Allowing for these extra currents, the interaction
is given by the Lagrangian
\begin{eqnarray}
{\cal L} = \sum_{V= \gamma, Z, W, \cdots} \ g_V J_V^{\mu} V_{\mu}
\end{eqnarray}

\nn where the complex conjugation of the charged currents is understood. The
currents $J_\mu^V$ can be expressed in terms of left--handed and right--handed
charges as
\begin{eqnarray}
J_\mu^{V}= \sum_{f} \overline{\psi}_f \gamma_\mu \left[\ Q_L^{ff'V} \ \frac{(1-
\gamma_5)}{2} \ + \ Q_R^{ff'V} \ \frac{(1+\gamma_5)}{2} \ \right] \psi_{f'}
\end{eqnarray}
For the minimal Standard Model gauge bosons, the coupling constants $g_V$ are
simply
\begin{eqnarray}
g_\gamma = e =\sqrt{4\pi \alpha} \ \ \ , \ \ \ \ g_Z=e/s_W c_W \ \ \ , \ \ \ \
g_W= e\sqrt{2}/s_W
\end{eqnarray}
with $e$ the proton charge and $s_W^2=1-c_W^2 \equiv \sin^2 \theta_W$; the
coupings $Q_{L,R}^{ff'V}$ read
\begin{eqnarray}
Q_{L,R}^{ff \gamma} = e^f \ \ , \hspace*{0.7cm} Q_{L,R}^{ffZ} = I_{3L,3R}^f
-e^f s_W^2 \ \ , \hspace*{0.7cm} Q_{L,R}^{ff'W}= |I_{3L,3R}|
\end{eqnarray}
with $e^f$ the electric charge in units of $e$ and $I_{3L},I_{3R}$ the third
components of weak isospin. \s

\nn The inclusion of additional gauge bosons in the previous equations is
straightforward once their couplings to fermions are specified. \s

\nn The	new fermions will mix with the ordinary fermions which have the same
U(1)$_{\rm Q}$ and SU(3)$_{\rm C}$ quantum numbers \cite{R12,R13}. The mixing
will determine the decay systematics of the new leptons and quarks and allows
for their single production in association with light fermions. In principle,
one has to treat the three generations of fermions on the same footing; this,
leads to rather complicated mixing patterns. For instance in E$_6$, the mixing
in the general case would be described by $6 \times 6 $ non--diagonal matrices
in the quark and charged lepton sectors and by $15 \times 15$ matrices in the
neutrino sector \cite{R4}. The matrix elements will depend on the vacuum
expectation values of the Higgs fields and on arbitrary Yukawa couplings,
making a general analysis rather complicated\footnote{However, the
inter--generational mixing will induce at the tree level, flavor changing
neutral currents  which are severely constrained by existing data \cite{R4};
neglecting the latter allows for an enormous simplification: the mixing can be
parametrized by a few angles which can be treated as phenomenological
parameters.}. \s

\nn In order not to commit ourselves to any particular model, we will allow for
the mixing between different generations and treat the mixing angles as
phenomenological parameters. For instance, in the case of the interaction of
the electron and its associated neutrino with exotic charged and neutral heavy
leptons, the general Lagrangian describing the transitions between $e,\nu_e$
and the heavy leptons $N^k,E^k$ where $k$ is a generation index reads
\begin{eqnarray}
{\cal L} & = & \frac{1}{2} \sum_{k=1}^{3} \ \sum_{i=L,R} \ \left[ g_W \
\zeta_i^{\nu E^k W} \bar{\nu_e} \gamma_\mu E_i^k \ W^\mu \ + \ g_Z \ \zeta_i^{
eE^k Z} \bar{e} \gamma_\mu E_i^k \ Z^\mu \right] \ + \ {\rm h.c.} \non \\
         & + & \frac{1}{2} \sum_{k=1}^{3}\ \sum_{i=L,R}\ \left[ g_W \ \zeta_i^{
e N^k W} \bar{e} \gamma_\mu N_i^k \ W^\mu \ +\ g_Z \ \zeta_i^{\nu N^k Z}
\bar{\nu_e} \gamma_\mu N_i^k \ Z^\mu \right] \ + \ {\rm h.c.}
\end{eqnarray}
where we have allowed for both left--handed and right--handed mixing and
assumed small angles so that one can write $\sin\zeta_{L,R} \simeq
\zeta_{L,R}$.
{}From this Lagrangian, where $g_{W,Z}$ are given in eq.~(2.3), the charges
$Q^{fFV}_{L,R}$ as in eq.~(2.2) can be easily derived. The generalization to
the other light leptons and to quarks as well as to extra gauge bosons is
straightforward.

\newpage

\subsection*{3.~Production in e$^+$e$^-$ Collisions}

\renewcommand{\theequation}{3.\arabic{equation}}
\setcounter{equation}{0}

\vspace*{3mm}

{\bf 3.1 General Case} \\

\nn In this subsection, we give the most general expression of the differential
cross section for the production of two fermions with different masses, to
treat
pair and single production on the same footing, including the longitudinal
polarization of the initial $e^+/e^-$ beams and the polarization of the final
fermions. \s

\nn Consider the process where a pair of heavy fermions is produced in $\ee$
annihilation through gauge boson exchange
\beq
e^+(\ov{l},\ov{\xi})\ e^-(l,\xi) \ \lra \ \ov{F}(\ov{p},\ov{n})\ F(p,n)
\eeq
$\xi$, $\ov{\xi}$ denote the degrees of longitudinal polarization of the
initial electron and positron and $l$, $\ov{l}$ their momenta;  $p(\ov{p}$),
$n(\ov{n})$ and $m(\ov{m})$ are the four--momentum, spin vector and mass of the
final fermion $F(\ov{F})$.
The most general form of the differential cross section, d$\sigma$
can be written in terms of the scalar products of the spin four--vectors and
momenta of the particles, as
\begin{eqnarray}
{\rm d}\sigma = \frac{1}{2}\frac{N_c e^4}{(l.\ov{l})} (2\pi)^4 \delta^4(l+
\ov{l}-p-\ov{p})
\frac{{\rm d}^3p}{(2\pi)^32p^0}\frac{{\rm d}^3\ov{p}}{(2\pi)^32\ov{p}^0}
\left[ (1-\xi \ov{\xi}) {\cal A} + (\ov{\xi}- \xi) {\cal A}' \right]
\end{eqnarray}
\nn where $e$ is the proton charge, $N_c$ the color factor of the final
fermions and the squared amplitudes ${\cal A}$ and ${\cal A}'$ can be
expressed in terms of generalized charges \cite{R21} $Q_{1,2,3}$ and
$Q'_{1,2,3}$
\begin{eqnarray}
{\cal A} &=& (p.\ov{l}) (\ov{p}.l)(Q_1+Q_3)+(p.l)(\ov{p}.\ov{l})
          (Q_1-Q_3)+ m \ov{m} (l.\ov{l}) Q_2 \non \\
      &-& (n.l)[\ov{m} (p.\ov{l})Q_2'+m (\ov{p}.\ov{l})(Q_1'-Q_3')]
      +   (n.\ov{l})[\ov{m} (p.l)Q_2'+m(\ov{p}.l)(Q_1'+Q_3')] \non \\
      &-& (\ov{n}.l)[m (\ov{p}.\ov{l})Q_2'+\ov{m}(p.\ov{l})(Q_1'+Q_3')]
       +  (\ov{n}.\ov{l})[m (\ov{p}.l)Q_2'+\ov{m}(p.l)(Q_1'-Q_3')] \non \\
      &+& (n_.\ov{n})Q_2[(l.\ov{l})(p.\ov{p})-(p.l)(\ov{p}.\ov{l})-(p.\ov{l})
           (\ov{p}.l)]
       -  2 Q_2[(n.l)(\ov{n}.l)+(n.\ov{l})(\ov{n}.\ov{l})](l.\ov{l})  \non \\
      &-& (n.\ov{l})(\ov{n}.l)[Q_2((l.\ov{l})+(p.\ov{p})+
          (\ov{p}.l)+(p.\ov{l}))  +  m\ov{m}(Q_1+Q_3)] \non \\
      &-& (n.l)(\ov{n}.\ov{l})[Q_2((l.\ov{l})+(p.\ov{p})+
          (p.l)+(\ov{p}.\ov{l})) +  m\ov{m}(Q_1-Q_3)] \nonumber
\end{eqnarray}
\begin{eqnarray}
{\cal A'} \ = \ {\cal A} \left( \ Q_1 \leftrightarrow Q_1' \ , \ Q_2
\leftrightarrow	Q_2' \ , \ Q_3 \leftrightarrow Q_3' \ \right)
\end{eqnarray}
\nn In terms of	the helicity amplitudes	$Q_{ij}$ with $i,j=L,R$, the charges
$Q$ and $Q'$ are
\begin{eqnarray}
Q_{1} &=& \frac{1}{4} \left[ \;	|Q_{LL}|^{2} + |Q_{RR}|^{2} +
|Q_{RL}|^{2} +|Q_{LR}|^{2} \right]  \nonumber \\
Q_{2} &=& \frac{1}{2} {\rm Re} \ \left[	\; Q_{LL}Q_{RL}^{*} +
Q_{RR}Q_{LR}^{*} \right]  \nonumber \\
Q_{3} &=& \frac{1}{4} \left[ \;	|Q_{LL}|^{2} + |Q_{RR}|^{2} -
|Q_{RL}|^{2} - |Q_{LR}|^{2} \right]  \nonumber \\
Q_1' &=& \frac{1}{4} \left[ \;	|Q_{LL}|^{2} + |Q_{RL}|^{2} -
|Q_{RR}|^{2} -|Q_{LR}|^{2} \right]  \nonumber \\
Q_2' &=& \frac{1}{2} {\rm Re} \ \left[	\; Q_{LL}Q_{RL}^{*} -
Q_{RR}Q_{LR}^{*} \right]  \nonumber \\
Q_3' &=& \frac{1}{4} \left[\;	|Q_{LL}|^{2} + |Q_{LR}|^{2} -
|Q_{RR}|^{2} - |Q_{RL}|^{2} \right]
\end{eqnarray}

\nn The	helicity amplitudes $Q_{ij}$ depend on the process under consideration.
For $s=(l+\ov{l})^2$, $t=(p-l)^2$ and $u=(p-\ov{l})^2$  channel vector bosons
$V_S$, $V_T$ and $V_U$ exchange, respectively, the general form is
\begin{equation}
Q_{ij}=\sum_{V_S} \frac{g_{V_S}^2}{e^2} \frac{Q_i^{FFV_S}Q_j^{eeV_S} \ s }
{s-M_{V_S}^2 +i\Gamma_{V_S} M_{V_S}} + \sum_{V_T}\frac{g_{V_T}^2}{e^2}
\frac{Q_i^{eFV_T} Q_j^{eFV_T}s}{t-M_{V_T}^2} + \sum_{V_U}\frac{g_{V_U}^2}
{e^2} \frac{Q_i^{eFV_U}Q_j^{eFV_U}s}{u-M_{V_U}^2}
\end{equation}

\nn The normalization factors $g_{V}$ and the reduced couplings
$Q_{L,R}^{ff'V}$
can be derived from the Lagrangian describing the $ff'V$ interaction; see
section 2. \s

\nn Integrating over the variables of one of the final fermions as well as on
the azimuthal angle of the remaining one, the differential cross section
d$\sigma/ $d$\cos\theta$, where $\theta$ specifies the direction of the latter
particle with respect to the incoming electron, reads
\begin{eqnarray}
\frac{{\rm d} \sigma}{{\rm d} \cos \theta } = \frac{3}{8} \ \sigma_0 \ N_c
\lambda^{\frac{1}{2}} \  \frac{1}{4} \ \left[ \ (1-\xi \ov{\xi}){\cal A} +
(\ov{\xi} -\xi) {\cal A}' \ \right]
\end{eqnarray}

\nn where $\sigma_0=4\pi \alpha^2/3s$ is the point--like QED cross section for
muon pair production  and $\lambda$ the usual two body phase space function
\begin{eqnarray}
\lambda=(1-\mu^2 -\ov{\mu}^2)^2 -4\mu^2 \ov{\mu}^2 \ \ , \hspace*{0.7cm}
{\rm with} \ \  \mu=m/\sqrt{s} \ , \ \ov{\mu}=\ov{m}/\sqrt{s}
\end{eqnarray}
In terms of the charges $Q_i$ and $Q'_i$, the reduced amplitudes squared
${\cal A}$ and ${\cal A}'$ read
\begin{eqnarray}
{\cal A} &=& \left[ 1 -(\mu^2-\ov{\mu}^2)^2 + \lambda \cos^2 \theta
\right]	Q_1 + 4	\mu \ov{\mu} Q_2	+ 2 \lambda^{\frac{1}{2}} \cos\theta
Q_3 \non \\
& - & 2 \frac{m}{s} \ n \cdot (l-\ov{l}) \left[ (1 -\mu^2+\ov{\mu}^2)
Q_1' +(1+\mu^2 -\ov{\mu}^2) \frac{\ov{m}}{m} Q_2' + \lambda^{\frac{1}{2}}
\cos\theta Q_3'	\right]	\non \\
& +& 2 \frac{m}{s} \ n \cdot (l+\ov{l}) \left[ (1 -\mu^2+\ov{\mu}^2)Q_3'
+ \lambda^{\frac{1}{2}}	\cos\theta ( Q_1' - \frac{\ov{m}}{m} Q_2')
\right]	\non \\
& + & 2 \frac{\ov{m}}{s} \ \ov{n} \cdot (\ov{l}-l) \left[ (1 -\ov{\mu}^2+\mu^2)
Q_1' +(1+\ov{\mu}^2 -\mu^2) \frac{m}{\ov{m}} Q'_2 + \lambda^{\frac{1}{2}}
\cos\theta Q_3'	\right]	\non \\
& - & 2 \frac{\ov{m}}{s} \ \ov{n} \cdot (l+\ov{l}) \left[ (1 -\ov{\mu}^2+\mu^2)
Q_3'+ \lambda^{\frac{1}{2}}	\cos\theta ( Q_1' - \frac{m}{\ov{m}} Q_2')
\right]	\non \\
& +& n \cdot \ov{n} \ (1-\cos^2 \theta) \ \lambda \ Q_2 -\frac{8}{s}Q_2
[n\cdot l\ \ov{n}\cdot (l+\ov{l})+\ov{n}\cdot \ov{l}\ n \cdot (l+\ov{l})]\non
\\
&+& \frac{4}{s} n \cdot l \ \ov{n} \cdot \ov{l} \ [(1+\mu^2+\ov{\mu}^2+
\lambda^{\frac{1}{2}} \cos \theta)Q_2+ 2 \mu \ov{\mu} (Q_3-Q_1) ] \non \\
&-& \frac{4}{s} \ov{n} \cdot l \ n \cdot \ov{l} \ [(3-\mu^2-\ov{\mu}^2+
\lambda^{\frac{1}{2}} \cos \theta)Q_2+2 \mu \ov{\mu} (Q_3+Q_1) ] \nonumber
\end{eqnarray}
\begin{eqnarray}
{\cal A}' \ = \	{\cal A} \ \left( \ Q_1	\leftrightarrow	Q_1' \ , \ Q_2
\leftrightarrow	Q_2' \ , \ Q_3 \leftrightarrow Q_3' \ \right)
\end{eqnarray}

\vspace*{3mm}

\nn The	polarization four--vector $P_\mu$ of the final state fermion $F$ is
defined by d$\sigma^{{\rm pol}}/{\rm d}\cos \theta \sim {\rm d}\sigma^{{\rm
unpol}} / {\rm d} \cos \theta \times [1 + P_\mu n^\mu]$, with $n_\mu$ the spin
vector which satisfies the relations $n \cdot n=-1$ and $n \cdot p=0$; see
Appendix.  In the $F$ rest frame, assuming CP--conservation, the components
are $(0, P_\perp, 0, P_{||})$ with $P_\perp$ and $P_{||}$ the transverse and
longitudinal polarizations with respect to the flight direction. Summing over
the polarizations of one of the final fermions, e.g. $\ov{F}$, the longitudinal
and transverse components of the polarization vector of the other fermion, in
its own rest frame, are given by
{\small
\begin{eqnarray}
{\cal P}_{||} & = & \frac{[1 - \mu^2 + \ov{\mu}^2 +(1 + \mu^2 -
\ov{\mu}^2)\cos^2\theta] \lambda^{\frac{1}{2}}
Q_3' + \cos \theta [ (1+\mu^2-\ov{\mu}^2-\lambda)\frac{\ov{m}}{m} Q_2'
+2(1-\mu^2-\ov{\mu}^2)Q_1'] } {-\left[ 1 -(\mu^2-\ov{\mu^2})^2 + \lambda
\cos^2 \theta \right] Q_1 - 4 \mu \ov{\mu} Q_2 -	2 \lambda^{\frac{1}{2}}
\cos\theta Q_3}	\non \\
{\cal P}_\perp & =& 2 \mu \sin \theta \frac{(1 - \mu^2 + \ov{\mu}^2)
Q_1'
+(1 + \mu^2 - \ov{\mu}^2) \frac{\ov{m}}{m}Q_2' + \lambda^{\frac{1}{2}}
\cos \theta Q_3'} {\left[ 1-(\mu^2-\ov{\mu}^2)^2 + \lambda \cos^2 \theta
\right] Q_1 + 4 \mu \ov{\mu} Q_2 + 2 \lambda^{\frac{1}{2}} \cos\theta Q_3}
\end{eqnarray} }

\vspace*{2mm}

\nn In the next subsections, the special cases of heavy fermion pair production
and single production in association with massless fermions will be discussed.
\\
\vspace*{2mm}

\nn {\bf 3.2.~Pair Production} \\

\nn In $\ee$ collisions, the pair production of new fermions proceeds through
$s$--channel gauge boson exchange; there are also contributions from
$t$--channel exchange in the case of heavy lepton production, but they are
quadratically suppressed by mixing angle factors and therefore, rather small.
The unpolarized differential cross section ${\rm d}\sigma/{\rm d} \cos \theta$
for the process $\ee \ra F \bar{F}$ is (see also Ref.~\cite{R20})
\begin{eqnarray}
\frac{{\rm d} \sigma}{{\rm d} \cos \theta } = \frac{3}{8} \ \sigma_0
\ N_c \beta_F \left[ (1+\beta_F^2 \cos^2 \theta)Q_1+ (1-\beta^2_F)Q_2
+2  \beta_F  \cos \theta Q_3 \right]
\end{eqnarray}
\nn with $\beta_F=(1-4m_F^2/s)^{1/2}$ the velocity of the fermion in the final
state; the charges $Q_1,Q_2$ and $Q_3$ are given by in eq.~(3.4) with the
helicity amplitudes $Q_{ij}$ with $i,j=L,R$ in the general case in eq.~(3.5).
If only $s$--channel photon and Z boson exchange is present, these helicity
amplitudes  are simply given by
\begin{equation}
Q_{ij} \ = \ e^Fe^e \ +\ \frac{Q_i^{FFZ}Q_j^{eeZ}}{s_W^2c_W^2}
\frac{s}{s- M_Z^2+i\Gamma_Z M_Z}
\end{equation}
\nn where $Q^{ffZ}_{i}$ are the reduced couplings of the left and right--handed
fermions to the $Z$ boson. \s

\nn In the case of unpolarized initial beams, the cross section eq.~(3.10)
allows for three independent measurements: the total production cross section
$\sigma_F$, the forward--backward asymmetry $A^{FB}_F$ and the parameter
$\alpha_F$ defined as d$\sigma/$d$\cos \theta \sim 1+\alpha_F \cos^2 \theta$.
These three parameters, and their asymptotic values,  are given by

\begin{eqnarray}
\sigma_F = &  \frac{3}{4} \sigma_0 N_{c} \beta_F
\left[(1+\frac{1}{3}\beta_F^{2}
) Q_{1}+(1-\beta_F^{2})Q_{2} \right]  \; \; \; & \stackrel{ \sqrt{s} \gg m_{F}}
{\longrightarrow} \; \; N_{c} \sigma_0 Q_{1} \\
A^{FB}_F = & \beta_F Q_{3} \left[ (1+\frac{1}{3}\beta_F^{2})Q_{1}+
(1-\beta_F^{2})
Q_{2} \right]^{-1} \; \; \; & \stackrel{ \sqrt{s} \gg m_{F}} {\longrightarrow}
\; \; \frac{3}{4} \frac{Q_{3}}{Q_{1}}  \\
\alpha_F = &  \beta_F^2 Q_{1} \left[ Q_{1}+ (1-\beta_F^{2}) Q_{2} \right]^{-1}
\; \; \; & \stackrel{ \sqrt{s} \gg m_{F}} {\longrightarrow} \; \; 1
\end{eqnarray}

\nn For Majorana neutrinos one has to symmetrize eq.~(3.10) because of the two
identical particles in the final state. This symmetrization makes that the
Majorana neutrino has only axial--vector couplings so that the charges are $Q_2
=-Q_1$ and $Q_3=0$,  leading to the simple expression for the cross section
\begin{eqnarray}
\sigma_{N_{\rm maj}} = \frac{1}{2} \sigma_0 \beta^3_N \ \left[
|Q_{LL}|^2+ |Q_{RL}|^2 \right]
\end{eqnarray}
The total cross section is proportional to $\beta^3$ and thus, strongly
suppressed near threshold; the angular distribution behaves like
d$\sigma$/d$\cos \theta \sim 1+\cos^2 \theta$ and therefore, there is no
forward--backward asymmetry and the $\alpha$ parameter is equal to one. Note
that since the isosinglet Majorana neutrinos do not couple to the photon and Z
boson, they can be pair produced only through the exchange of an extra gauge
boson and thus, the helicity amplitudes $Q_{ij}$ have to be altered.   \s

\nn Let us now discuss the polarization of the heavy fermions. Summing over
the polarizations of $\bar{F}$, the two components of the polarization vector
of $F$ in its own rest frame, $P_{||}$ and $P_\perp$ can be written as
(see also refs.~\cite{R20,R22})
\begin{eqnarray}
P_{||} & = & - \frac{(1+\cos^2\theta)\beta_F Q_3'+\cos \theta \left[
(1-\beta^2_F) Q_2'+ (1+\beta^2_F)Q_1' \right]}{ (1+\beta^2_F \cos^2\theta)Q_1
+(1-\beta^2_F)Q_2 +2 \beta_F \cos \theta Q_3 } \non \\
P_{\perp} & = & \sqrt{1-\beta^2_F} \sin \theta \frac{Q_1'+Q_2'+\beta_F \cos
\theta Q_3' }{ (1+\beta^2_F \cos^2\theta)Q_1+(1-\beta^2_F)Q_2+2 \beta_F \cos
\theta Q_3 }
\end{eqnarray}
\nn where the helicity amplitudes $Q_{ij}$ are given in eq.~(3.5) and (3.10)
and the charges  $Q'_{1,2,3}$ in eq.~(3.4). Averaged over the polar angle, the
two components become
\begin{eqnarray}
<P_{||}> & = & - \frac{4}{3} \frac{\beta_F Q_3'} {(1+\beta_F^2/3)Q_1+
(1-\beta^2_F)Q_2} \non \\
<P_{\perp}> & = & \frac{3\pi}{4} \frac{m_F}{\sqrt{s}} \frac{Q_1'+Q_2'} {(1+
\beta_F^2/3)Q_1+ (1-\beta^2_F)Q_2}
\end{eqnarray}

\nn Again, for Majorana neutrinos one has to symmetrize the previous
expressions. This makes that the currents become purely axial-vector, and
therefore there is no polarization effect. The polarization of the final
particles, together with their angular distributions, permits to discriminate
between Majorana and Dirac neutrinos, or between particles with different
couplings [standard, mirror or vector couplings].

\newpage

\nn {\bf 4.~2.~Single Production} \\

\nn In $\ee$ collisions one can also have access to the new fermions via single
production in association with their light partners if fermion mixing is not
too small. Assuming that extra gauge bosons are too heavy to affect the
production, the process proceeds through $s$--channel $Z$ exchange for all
fermions, but for heavy leptons [only the first family if
inter--generational mixing is neglected] one has additional $t$--channel
gauge boson exchanges: $W$ exchange for neutral leptons and $Z$ exchange
for charged leptons. Neglecting the mass of the light fermion partner, the
differential cross section for the process $\ee \ra F \bar{f}$ is
\begin{eqnarray}
\frac{{\rm d} \sigma}{{\rm d} \cos \theta } = \frac{3}{8} \ \sigma_0
\ N_c (1-\mu^2)^2 \left[ \left( 1+ \mu^2+ (1-\mu^2) \cos^2 \theta \right)Q_1
+2  \cos \theta Q_3 \right]
\end{eqnarray}
\nn where $\mu^2=m_F^2/s$. The charges $Q_{1,3}$ as given in eq.~(3.5)
are built--up by the helicity amplitudes
\begin{eqnarray}
Q_{ij} &=& \frac{1}{2} \frac{\zeta_i^{eE^kZ} Q^{eeZ}_j}{s_W^2c_W^2} \left[ \
\frac{1}{1-z} +\frac{1}{t/s-z} \ \right] \hspace{1.8cm} {\rm for} \ \ E^k
\non \\
Q_{ij} &=& \frac{\zeta_i^{\nu N^k Z} Q^{eeZ}_j}{2s_W^2c_W^2} \frac{1}{1-z}
+ \frac{\zeta_i^{eN^k W} Q_j^{e \nu W}} {s_W^2} \frac{1}{t/s-w} \hspace{0.6cm}
{\rm for} \ \ N^k \non \\
Q_{ij} &=& \frac{1}{2} \frac{ \zeta_i^{fFZ} Q^{eeZ}_j}{s_W^2c_W^2}
\frac{1}{1-z} \hspace*{4.6cm} {\rm for} \ \ {\rm quarks}
\end{eqnarray}
with the reduced masses $z=M_Z^2/s$, $w=M_W^2/s$ and $t/s=-(1-\cos \theta)(1-
\mu^2)/2$. Note again that in the case of Majorana neutrinos, the $Q_{ij}$
have to be symmetrized. \s

\nn For quarks [and for second and third family leptons if inter--generational
mixing is neglected] the total cross section $\sigma (\ee \ra F\bar{f}$) takes
the simple form
\begin{eqnarray}
\sigma (\ee \ra F\bar{f}) =  \sigma_0 \ N_{c} (1-\mu^2)^2 \left( 1+
\frac{1}{2} \mu^2 \right) Q_1
\end{eqnarray}

\nn The cross sections for the production of the conjugate states $\sigma
(\ee \ra \bar{F}f)$ is the same. For first generation heavy leptons, the
analytical expressions are much more involved because of the $t$--channel
contributions. Denoting by $V_S$ the gauge boson exchanged in the $s$--channel
[$Z$ for both types of leptons] and by $V_T$ the one exchanged in the
$t$--channel [$W$ for $N$ and $Z$ for $E$], one can write a common expression
for heavy lepton $L=N^k, E^k$ single production
\begin{eqnarray}
\sigma && \hspace*{-0.9cm} (\ee \ra L \bar{l}) \ = \ 3\sigma_0 \ \left\{ \
\frac{1}{3}(1-\mu^2)^2 \ \left[1+\frac{\mu^2}{2}\right] \ \frac{q_+^{V_SV_S}+
q_-^{V_SV_S}}{(1-v_S)^2} \right. \non \\
&+& \left[ (1-\mu^2)(3+2v_T-\mu^2)-2(1-\mu^2+v_T)(1+v_T)\ \La \right]
\frac{q_-^{V_SV_T}} {1-v_S} \non \\
&+&  \left[ -(1-\mu^2)(1+\mu^2-2v_T)-2v_T(\mu^2-v_T) \ \La \right]
\frac{q_+^{V_SV_T}}{1-v_S} \non \\
&+&  \left[ (1-\mu^2) \frac{1+2v_T}{v_T} - (2-\mu^2+2v_T)\ \La \right]
q_-^{V_TV_T} \non \\
&+&  \left. \left[ (1-\mu^2) \left( 2 -\frac{1}{1+v_T-\mu^2} \right) -(2v_T
-\mu^2) \ \La \right] q_+^{V_TV_T} \right\}
\end{eqnarray}

\nn where $v_S=M_{V_S}^2/s$, $v_T=M_{V_T}^2/s$ and the charges $q_\pm^{VV'}
=q_\pm^{V_S  V_T} \cdots$, are defined by
\begin{eqnarray}
q_+^{VV'} =  \frac{1}{4} \left[ |q_{LL}^{VV'}|^2 + |q_{RR}^{VV'}|^2 \right]
\hspace{2cm}
q_-^{VV'} = \frac{1}{4} \left[ |q_{LR}^{VV'}|^2 + |q_{RL}^{VV'}|^2 \right]
\non
\end{eqnarray}
with $q_{ij}^{V_SV_S} = (g_{V_S}^2/e^2) Q_i^{LlV_S}Q_j^{eeV_S}$ $etc.$
\s

\nn To obtain the cross section for the production of $E$ and $N$ in the case
of left--handed or right--handed mixing is then straightforward: one has simply
to specify $V_S$ and $V_T$ and choose the proper combination of charges. For
instance, in the case of the production of a heavy neutrino with a left--handed
mixing denoted\footnote{In principle, the indices $L,R$ refer to the handedness
of the heavy lepton mixing with its  light partner. However, due to the fact
that the latter is massless, they are also the chirality of the heavy lepton.}
$N_L$, one recovers up to a factor of two, the formula given in Ref.~\cite{R18}
for the production of Majorana neutrinos, by setting
\begin{eqnarray}
v_S=\frac{M_Z^2}{s} \ \ , \hspace{1cm} v_T= \frac{M_W^2}{s} \ \ , \hspace{1cm}
q_-^{WW}=q_-^{WZ}=0
\end{eqnarray}
\nn This is due to the fact that the production cross section of a Majorana
neutrino is just the sum of the production cross sections of a Dirac neutrino
and its anti--neutrino; this also holds true for the angular distributions. \s

\nn For $E$ production, we have a $Z$ exchange in both $s$ and $t$--channels.
This simplifies considerably the expression eq.~(3.21) since in addition to the
fact that $v_S=v_T$,  the charges factorize. Note that for $s_W^2 =1/4$, the
left--handed and right--handed couplings of the electron to the $Z$ are equal
in magnitude but with opposite signs. This leads to
\begin{eqnarray}
q_-^{ZZ} \simeq 0 \ \ , \hspace{1cm} q_+^{ZZ} \ \simeq \ \frac{1}{2}
|q_{LL}^{ZZ}|^2 \ \simeq \ \frac{1}{2} |q_{RR}^{ZZ}|^2
\end{eqnarray}
\nn which translates into the fact the cross sections for $E_L$ and $E_R$
are approximately equal. \\

\nn Finally, the longitudinal and transverse components of the polarization
vector of the heavy fermion produced in association with a massless partner
reads
\begin{eqnarray}
P_{||} & = & - \frac{[1-\mu^2+(1+\mu^2) \cos^2 \theta]Q_3' + 2
\cos \theta Q_1' }{ \left[ 1 +\mu^2 + (1-\mu^2) \cos^2 \theta \right] Q_1 +
2 \cos\theta Q_3}\non \\
P_\perp & =& 2 \mu \sin \theta \frac{ Q_1' + \cos \theta Q_3'}
{\left[ 1 +\mu^2 + (1-\mu^2) \cos^2 \theta \right] Q_1 + 2 \cos\theta Q_3}
\end{eqnarray}
where the charges $Q_{1,2,3}'$ are given in eq.~(3.4) with the helicity
amplitudes eq.~(3.19).

\newpage

\renewcommand{\theequation}{4.\arabic{equation}}
\setcounter{equation}{0}

\subsection*{4. Decays and Correlations}

\vspace*{2mm}

\nn {\bf 4.1 Two--body Decays} \\

\nn The heavy heavy fermion $F$, with a mass $m$ and spin four--vector $n$,
will decay into a lighter fermion $f_0$ and virtual or real gauge bosons
which subsequently decay into two massless fermions $f_1$ and $\bar{f}_2$
\beq
F(p ,n )   \ \longrightarrow \ f_0(l_0) \ V(l_V) \ \longrightarrow \ f_0(l_0)
\ f_1(l_1) \ \ov{f}_2(l_2)
\eeq
In most of the cases the fermion $f_0$ is just the ordinary partner of $F$
[which can be considered as massless] and the decay occurs through mixing.
But it is possible that $f_0$ is also a heavy fermion and the process eq.~(4.1)
 is a ``cascade" decay. For instance, for fermions belonging to the same
isodoublets, the heavier fermion can decay through the exchange of a $W$ boson
into its lighter isospin partner and the latter subsequently decays through
mixing into three massless particles. Moreover, it is also possible that
the mixing between heavy fermions is much larger than the mixing between the
new and ordinary fermions in which case the decay of the heavy fermion first
into a lighter one, is more important although kinematically disfavored. \s

\nn If the mass difference between $F$ and $f_0$ is larger than the mass of the
exchanged gauge boson, the latter will be on mass--shell and the decay is a
two--body decay. Assuming that $f_0$ is also polarized [the mass and spin
four--vector will be denoted by $m_0$ and $n_0$ respectively], the differential
decay width in terms of the momenta and spin--vectors of $F$ and $f_0$ is given
by
\beq
{\rm d}\Gamma = (2\pi)^4 \delta^4 (p- l_0-l_V) \ \frac{{\rm d}^3l_0}
{(2\pi)^32l_0^0} \ \frac{{\rm d}^3l_V}{(2\pi)^32l_V^0} \ \frac{e^2}{2m}
\ {\rm d}\Gamma^0
\eeq
with
\beq
{\rm d}\Gamma^0 &=& \left[ m^2+m_0^2-2M_V^2 +\frac{(m^2-m_0^2)^2}{M_V^2}
\right]
q_1 \ - \ 6m_0 m q_2 \non \\
&+& (n\cdot n_0) \left[ \left( m^2+m_0^2 -\frac{(m^2-m_0^2)^2}{M_V^2} \right)
q_2 +2 m m_0 q_1 \right] +2 (n_0 \cdot p)(n \cdot l_0) \frac{(m - m_0)^2}
{M_V^2} q_2 \non \\
&+& 2 (n\cdot l_0) \frac{m}{M_V^2} (m^2-m_0^2-2M_V^2) q_3
   +2 (n_0 \cdot p) \frac{m_0}{M_V^2} (m_0^2-m^2-2M_V^2) q_3
\eeq
In terms of the left and right--handed $Ff_0V$ couplings, the charges
$q_{1,2,3}$ are
\beq
q_{1,3} =\frac{1}{4} \frac{g_V^2}{e^2} \left[ (Q_L^{Ff_0V})^2 \pm  (Q_R^{Ff_0
V})^2 \right] \hspace*{0.4cm} , \hspace*{0.8cm}
q_2 =\frac{1}{2} \frac{g_V^2}{e^2} Q_L^{Ff_0V}Q_R^{Ff_0V}
\eeq

\nn Summing over the polarization of $f_0$, the angular distribution
d$\Gamma/$d$\cos\theta_0$ where $\theta_0$ is the angle between the spin vector
$n$ and the flight direction of $f_0$, can be written as
\begin{eqnarray}
\frac{{\rm d}\Gamma}{ {\rm d} \cos \theta_0} = \frac{1}{2} \left( \Gamma_{\rm
tot} \ + \ \cos \theta_0 \Gamma_{\rm ang} \right)
\end{eqnarray}

\nn $\Gamma_{\rm tot}$ is obtained by integrating the previous expression over
the angle $ \theta_0$
\begin{eqnarray}
\Gamma_{\rm tot} \ = \ \int_{-1}^{+1} \ {\rm d} \cos \theta_0 \ \frac{ {\rm d}
\Gamma}{ {\rm d} \cos \theta_0 }
\end{eqnarray}
and corresponds to the partial decay width for on--shell gauge bosons.
The $\cos \theta_0$ term can be isolated by integrating d$\Gamma$/d$ \cos
\theta_0$ asymmetrically
\begin{eqnarray}
\Gamma_{\rm ang} \ = \ \int_{0}^{+1} \ {\rm d} \cos \theta_0 \ \frac{ {\rm d}
\Gamma}{ {\rm d} \cos \theta_0 } \ -
\ \int^{0}_{-1} \ {\rm d} \cos \theta_0 \ \frac{ {\rm d} \Gamma}
{ {\rm d} \cos \theta_0}
\end{eqnarray}

\nn The expressions of $\Gamma_{\rm tot}$ and $\Gamma_{\rm ang}$ are
\beq
\Gamma_{\rm tot} &=& \frac{\alpha}{2} \frac{m^3}{M_V^2}\lambda^{\frac{1}{2}}
\left\{ \left[ (1-\mu_0^2)^2+\mu_V^2(1+\mu_0^2-2\mu_V^2)
\right] q_1 -6 \mu_0 \mu_V^2 q_2 \right\} \\
\Gamma_{\rm ang} &=& \frac{\alpha}{2} \frac{m^3}{M_V^2}
\lambda  (1-\mu_0^2-2\mu_V^2) q_3
\eeq
where
\beq
\lambda= (1-\mu_0^2-\mu_V)^2 -4\mu_0^2 \mu_V^2 \ \ \ \ , \ \ \ \
{\rm with} \ \mu_0= m_0/m \  , \  \mu_V = M_V/m
\eeq
The partial decay width and the angular distribution in the case where the
heavy fermion directly decays into its light partner and a gauge boson $V$ can
be obtained from the last expression by simply setting $m_0=0$ in the previous
expressions; one has
\beq
\Gamma_{\rm tot} &=& \frac{\alpha}{8} \frac{g_V^2}{e^2} \frac{m^3}{M_V^2}
(1-\mu_V^2)^2 (1+ 2\mu_V^2) \left[ (Q_L^{Ff_0V})^2 + (Q_R^{Ff_0
V})^2 \right] \\
\Gamma_{\rm ang} &=& \frac{\alpha}{8} \frac{g_V^2}{e^2} \frac{m^3}{M_V^2}
(1-\mu_V^2)^2 (1-2\mu_V^2)  \left[ (Q_L^{Ff_0V})^2 - (Q_R^{Ff_0V})^2 \right]
\eeq
in agreement with Ref.~\cite{R20}, once the charges are specified. \\

\nn In the next subsection, we will discuss the case where the exchanged gauge
bosons are off--shell, leading to three--body decays of the heavy fermions. \\

\nn {\bf 4.2~Three Body Decays} \\

\vspace*{2mm}

\nn The amplitude for the decay, eq.~(4.1), in the general case where the
fermion $f_0$ is also massive and polarized can be obtained from the amplitude
eq.(3.3) of the production of two heavy fermions by crossing symmetry: one
simply has to change the labels of the four--momenta and to reverse the sign of
the mass $\ov{m}$. One obtains for the differential decay width
\begin{eqnarray}
{\rm d}\Gamma = (2\pi)^4 \delta^4(p- l_0-l_1-l_2) \frac{{\rm d}^3l_0}
{(2\pi)^32l_0^0} \frac{{\rm d}^3l_1}{(2\pi)^32l_1^0} \frac{{\rm d}^3l_2}
{(2\pi)^32l_2^0} \ \frac{8N_c e^4}{m^5} \ {\rm d}\Gamma^0
\end{eqnarray}
where, after some simplifications, ${\rm d}\Gamma^0$ reads
\begin{eqnarray}
2 {\rm d}\Gamma^0 &=& (p.l_2)(l_0.l_1)(Q_1+Q_3)+
          (p.l_1)(l_0.l_2) (Q_1-Q_3)- mm_0 (l_1.l_2) Q_2 \non \\
      &+& (n.l_1)[m_0 (p.l_2)Q_2'- m (l_0.l_2) (Q_1'-Q_3')]
       -  (n.l_2)[m_0 (p.l_1) Q_2'- m (l_0.l_1) (Q_1'+Q_3')] \non \\
      &-& (n_0.l_1) [m (l_0.l_2) Q_2'- m_0 (p.l_2) (Q_1'+Q_3')]
       +  (n_0.l_2) [m (l_0.l_1) Q_2'- m_0 (p.l_1) (Q_1'-Q_3')] \non \\
      &+& (n.l_2)(n_0.l_1) [ m^2Q_2 +  mm_0 (Q_1+Q_3)] - (n.l_1)(n_0.l_2)
           [m^2 Q_2+  m m_0(Q_1-Q_3)] \non \\
      &+& (n.n_0) [ (l_1.l_2)(p.l_0) -(p.l_1)(l_0.l_2)-(p.l_2)(l_0.l_1)]  Q_2
\non \\
&-& 2 [(n.l_1)(n_0.l_2) + (n.l_2)(n_0.l_2) ](l_1. l_2) Q_2
\end{eqnarray}
\nn The generalized charges $Q_i$ and $Q'_i$ are the same as those given for
the production amplitudes, eq.~(3.4), but they are built--up with the helicity
amplitude
\begin{equation}
Q_{ij} = \frac{g_V^2}{e^2} Q_i^{Ff_0V}Q_j^{f_1f_2V} \ \frac{m^2}
{(l_1+l_2)^2- M_V^2 +i\Gamma_V M_V}
\end{equation}
assuming that the decay is mediated by only one gauge boson exchange; other
channels can be easily included. \\

\nn In the case where the fermion $f_0$ is also massless, as it happens in most
of the cases: the heavy fermion directly decays through mixing into its light
partners and real or virtual gauge bosons which subsequently decay into two
massless fermions, one can sum over its polarization and the expression
${\rm d}\Gamma^0$ simplifies considerably
\begin{eqnarray}
{\rm d}\Gamma^0 &=& (p \cdot l_2)(l_0 \cdot l_1)(Q_1+Q_3)+(p\cdot l_1)
(l_0 \cdot l_2) (Q_1-Q_3) \non \\
      & & - m (n \cdot l_1)(l_0 \cdot l_2) (Q_1'-Q_3') + m (n\cdot l_2)
(l_0 \cdot l_1)(Q_1'+Q_3')
\end{eqnarray}

\nn However, since in this case several decay channels are possible, the
helicity amplitudes are involved and one has in the general case where
$s=(p-l_0)^2,~t=(p-l_1)^2$ and $u=(p-l_2)^2$ channels $V_S, V_T$ and $V_U$
exchanges are present\footnote{An example of a situation where all the three
channels occur is the decay of a heavy Majorana neutrino into an $e^+e^-$ pair
and a neutrino or antineutrino. There is an ``$s$--channel" $Z$--boson exhange
$N \ra \nu_e Z \ra \nu_e \ee$, a ``t--channel" $W$--boson exchange $N \ra
e^-W^+ \ra \ee \nu_e$ and a ``u--channel" exchange $N \ra e^+W^- \ra \ee
\bar{\nu}_e$; if the light neutrino is also a Majorana particle, the three
amplitudes and the one due to $N \ra \bar{\nu}_e Z \ra \bar{\nu}_e \ee$ add
coherently.}

\begin{equation}
Q_{ij}= \sum_{V_S} \frac{g_{V_S}^2}{e^2} \frac{Q_i^{f_1f_2V_S}Q_j^{f_0FV_S} \
m^2 }{s-M_{V_S}^2 } + \sum_{V_T}\frac{g_{V_T}^2}{e^2}
\frac{Q_i^{f_0f_2V_T} Q_j^{f_1FV_T} m^2}{t-M_{V_T}^2
} + \sum_{V_U}\frac{g_{V_U}^2}{e^2} \frac{Q_i^{f_0f_1V_U}Q_j^{f_2FV_U}m^2}
{u-M_{V_U}^2}
\end{equation}

\vspace*{3mm}

\nn where the widths of the gauge bosons have been omitted for simplicity.
These expressions may need to be supplemented by statistical factors. \s

\nn However, in most physical situations there are at most two decay
``channels"
only, although in the same channel, several gauge bosons can be exchanged.
These two decay channels occur for instance, when two identitical particles
are present in the final state [e.g. $E \ra e^- Z \ra e^- \ee$] hence requiring
the symmetrization of the amplitudes or for decays involving both the isospin
up and down  light partners of the heavy fermions, leading to both neutral and
charged gauge boson exchanges [e.g. $E \ra e^- Z+\nu_eW^- \ra e^-\nu_e \bar{
\nu_e}$]. In the rest of the discussion, we will therefore assume that only two
channels are present and for simplicity, that there is only one gauge boson
exchanged in each channel.\s

\nn Integrating over the momentum of one of the final fermions, e.g.~$f_2$, as
well as on the azimuthal angle dependence, and using the usual scaled variables
\begin{eqnarray}
x_1=\frac{2\ (p \cdot l_1)}{m^2} \ \ , \ \ x_2=\frac{2\ (p \cdot l_2)}{m^2}
\ \ ,  \ \   x_0=\frac{2\ (p \cdot l_0)}{m^2} = 2-x_1-x_2
\end{eqnarray}
\nn the differential decay width writes
\begin{eqnarray}
\frac{ {\rm d} \Gamma} {  {\rm dcos}\theta_1 {\rm d}x_1 {\rm d} x_2} \ = \
N_c \frac{\alpha^2}{8\pi} m \  {\rm d}\Gamma_0
\end{eqnarray}
with $\theta_1$ the angle between the fermion $f_1$ and the spin vector $n$
of the heavy fermion [the projections of $n$ onto the flight directions of
$f_1$ and $f_2$ are simply $2n\cdot l_1/m = x_1 \cos \theta_1$ and $2n \cdot
l_2 /m =-x_2 \cos\theta_1$]; neglecting the widths of the exchanged gauge
bosons and using the scaled masses $v_S=M_{V_S}^2/m^2$ and $v_T=M_{V_T}^2/m^2$,
${\rm d}\Gamma_0$ is given by
\begin{eqnarray}
{\rm d}\Gamma_0 & = & \frac{x_1(1-x_1)q_{-}^{SS}+
x_2(1-x_2)q_{+}^{SS} } {(1-x_0-v_S)^2}
+ \frac{x_0(1-x_0) q_{-}^{TT} +x_2(1-x_2)q_+^{TT}}{(1-x_1-v_T)^2} \non \\
& - & \left[ \frac{ x_1(1-x_1)q_{-}^{'SS}+ x_2 (1-x_2) q_{+}^{'SS} }
{(1-x_0-v_S)^2} +\frac{x_1(1-x_0)q_+^{'TT} + x_2(1-x_2)q_+^{'TT}}{(1-x_1-
v_T)^2} \right] \cos\theta_1 \non \\
&-& 2 \ \eta \ q_+^{ST} \ \frac{x_2(1-x_2)}{(1-x_0-v_S)(1-x_1-v_T)}
\end{eqnarray}

\nn Here,  $\eta$ is a statistical factor: $\eta=-1$ in the case where there
are
two identical Dirac fermions in the final state otherwise $\eta=+1$; for quarks
one has also to divide by the color factor, i.e.~$\eta=\pm1/3$. The charges
$q_{\pm}^{MN}$ and $q_{\pm}^{'MN}$ where $M,N=S,T$ describe the channels with
$V_S, V_T$ exchange and the interference term; they read
\begin{eqnarray}
q_+^{MN~} = \frac{1}{2} \left[ |q_{LL}^{M}\ q_{LL}^{N} |
+ |q_{RR}^{M} \ q_{RR}^{N} | \right] \ \ \ & , & \hspace*{0.5cm}
q_-^{MN~} = \frac{1}{2} \left[ |q_{RL}^{M} \ q_{RL}^{N} |
+ |q_{LR}^{M} \ q_{LR}^{N} | \right] \non \\
q_+^{'MN} = \frac{1}{2} \left[ |q_{LL}^{M} \ q_{LL}^{N} |
- |q_{RR}^{M} \ q_{RR}^{N} | \right]  \ \ \ & , & \hspace*{0.5cm}
q_-^{'MN} = \frac{1}{2} \left[ |q_{RL}^{M} \ q_{RL}^{N} |
- |q_{LR}^{M} \ q_{LR}^{N} | \right]
\end{eqnarray}
where the helicity amplitudes $q_{ij}^{M}$ with $i,j=L,R$ are given by
\begin{eqnarray}
q_{ij}^{S} = \frac{g_{V_S}^2}{e^2} Q_i^{F f_0V_S}Q_j^{f_1f_2V_S} \
\hspace*{0.5cm} , \hspace*{1cm}
q_{ij}^{T} = \frac{g_{V_T}^2}{e^2} Q_i^{F f_1V_T}Q_j^{f_0f_2V_T} \
\end{eqnarray}

\nn Integrating over the energy of the particle $f_2$ with the boundaries
$1-x_1 \leq x_2 \leq 1$, one has
\begin{eqnarray}
\frac{{\rm d}\Gamma}{ {\rm d}x_1 {\rm d} \cos \theta_1} &=&
N_c \frac{\alpha^2}{8\pi} m \left\{
\frac{x_1^2(1-x_1)}{v_S(v_S-x_1)} \left[q_-^{SS}-q_-^{'SS}\cos \theta_1 \right]
+\frac{x_1^2(1-5x_1/3)}{2(1-x_1-v_T)^2} q_-^{'TT} \cos \theta_1  \right. \non
\\
&+& \left[ (2x_1-2v_S-1) \log \frac{v_S-x_1}{v_S} +\frac{x_1}{v_S} (x_1-1-
2v_S) \right] \left[q_+^{SS}- q_+^{'SS}\cos \theta_1 \right] \non \\
&+& \frac{x_1^2(1-2x_1/3)}{2(1-x_1-v_T)^2} \left[ q_+^{TT} - q_+^{'TT} \cos
\theta_1 +q_-^{TT} -q_-^{'TT} \cos \theta_1 \right] + \frac{2 \ \eta q_+^{ST}}
{1-x_1-v_T} \non \\
&\times & \left. \left[ (1-x_1+v_S)(v_S-x_1) \log\frac{v_S-x_1}{v_S}
- x_1 \left( \frac{3}{2}x_1-1-v_S \right) \right] \right\}
\end{eqnarray}

\nn Finally, when the energy of $f_1$ is also integrated out, one has for the
decay distribution
\begin{eqnarray}
\frac{{\rm d}\Gamma}{ {\rm d} \cos \theta_1} = \frac{1}{2} \left( \Gamma_{\rm
tot} \ + \ \cos \theta_1 \Gamma_{\rm ang} \right)
\end{eqnarray}
where $\Gamma_{\rm tot}$ corresponds to the partial decay width for off--shell
intermediate vector bosons and $\Gamma_{\rm ang}$ to the angular distribution.
They are given by
\begin{eqnarray}
\Gamma_{\rm tot} &=& N_c \frac{\alpha^2}{8\pi} m \left[ \left( q_-^{SS}
+ q_+^{SS} \right) R_T(v_S) + \left( q_-^{TT} + q_+^{TT} \right)  R_T(v_T)
+ \eta q_+^{ST} R_I(v_S,v_T) \right]  \hspace*{5mm} \\
\Gamma_{\rm ang} &=& - N_c \frac{\alpha^2}{8\pi} m \left[ \left( q_-^{'SS}
+ q_+^{'SS} \right) R_T(v_S) + \left( q_-^{'TT} + q_+^{'TT} \right)  R_T(v_T)
+ q_+^{'TT} R_A(v_T) \right]
\end{eqnarray}
where $R_T(v)$ describes the decay width in a given channel,
\begin{eqnarray}
R_T (v) = v(v-1) \log \frac{v-1}{v}+ \frac{1}{6v} (6v^2-3v-1)
\end{eqnarray}
$R_I(v_S,v_T)$ describes the interference between the two channels when
simultaneoulsy present. In terms of the Spence funtion, Li$_2(x)=-\int_0^1
dy~y^{-1} \log (1-xy)$, it is given by
\begin{eqnarray}
R_I (v,v) &=& (1-v_T)(1-2v_S-3v_T) \log \frac{v_T}{v_T-1}
+ (1-v_S)(1-2v_T-3v_S)  \non \\
&\times & \log \frac{v_S}{v_S-1}+ 2(v_S+ v_T)(1-v_S-v_T) \left[ \log \frac{v_T}
{v_T-1} \log \frac{v_S+v_T-1}{v_S} \right. \non \\
&+& \left. {\rm Li}_2 \left( \frac{v_T-1}{v_S+v_T-1} \right)
- {\rm Li}_2 \left( \frac{v_T}{v_T+v_S-1} \right)  \right] +3-5v_S-5v_T
\end{eqnarray}
and $R_A$ describes, for only $V_T$ exchange, the deviation of the angular
distribution from the familiar d$\Gamma/$d$\cos \theta_1 \sim  (1-\cos\theta_1
)$ form,
\begin{eqnarray}
R_A (v) = \frac{1}{2}(v-1)(3-5v) \log \frac{v-1}{v}- \frac{1}{12v} (6v^2-21v-8)
\end{eqnarray}

\nn Once the charges are specified, the expressions of $R_V$ and $R_I$ agree
with those obtained in Ref.~\cite{R19} and the expression of $R_A$ [and $R_V$]
with the result of Ref.~\cite{R20}.

\newpage

\subsection*{5. Summary}

In this paper, we have analyzed in detail the decay modes and the production
mechanisms in $\ee$ collisions, of new heavy fermions predicted by extensions
of the Standard Model and in particular by Grand Unified Theories. \s

\nn We have given analytical expressions for the production of two heavy
fermions of any flavor in a general case: we have allowed for arbitrary
couplings of the new fermions, the presence of several production channels and
we have taken into account the polarization of the initial $\ee$ state and the
final polarization of the heavy fermions. We have treated the case where the
two fermions have different masses, to discuss the pair production and the
single production in association with ordinary fermions on the same footing,
and to account for the possibility of producing two different heavy leptons.
Complete and compact formulae were given for angular distributions, total cross
sections and polarization vectors. \s

\nn We have then discussed in detail the decay modes of the new fermions
including the cascade decays and the three--body decays with off--shell gauge
bosons. Complete analytical expressions were given for total widths as well as
for angular and energy distributions. We have also taken into account the
polarization  of the decaying fermion; combined with the spin--dependent cross
sections, the spin--dependent decay distributions allow to obtain the full
correlations between all the particles involved in the process. These
correlations are very useful to discriminate between different types of
particles and will be very important when discussing the signals for heavy
fermion production since they help to optimize the experimental cuts which
permit to suppress the various backgrounds without affecting drastically the
signal cross sections. \s

\nn More phenomenological aspects of heavy fermion production at future
high--energy $\ee$ linear colliders, including a detailed analysis of the
various signals and backgrounds, will be presented in a subsequent paper
\cite{R23}. \\

\vspace*{5mm}

\nn {\bf Aknowledgements.} \s

\nn We thank G.~Azuelos, F. Boudjema, W. Buchm\"uller, C. Greub, M. Spira
and P. Zerwas for various discussions. Special thanks go to Peter Zerwas, for
useful conversations on polarization. This work is supported in part by the
National Sciences and Engineering Research Council of Canada.

\newpage

\subsection*{Appendix: Correlated Production and Decay}

\vspace*{2mm}

\renewcommand{\theequation}{A.\arabic{equation}}
\setcounter{equation}{0}

In this Appendix, we briefly summarize the formalism introduced by Tsai
\cite{R15} to describe the production of heavy fermions in $\ee$ collisions
\beq
\ee \ \lra \ \ov{F} \ F
\eeq
which subsequently decay through real or virtual gauge boson exchange, into
three body massless final states
\beq
F & \longrightarrow & f_0 \ f_1 \ \ov{f}_2 \non \\
\ov{F} & \longrightarrow & \ov{f_0} \ \ov{f_1} \ f_2
\eeq
including the spin correlations. This formalism allows an easy reconstruction
of the correlation between the initial $e^+,e^-$ states and the final decay
products of the heavy fermions. \\

\nn The polarization vector $\cp(\ov{\cp})$ of the fermion $F(\ov{F})$ is
defined with the help its spin four--vector $n(\ov{n})$. The latter satisfies
the relations $n \cdot n = \ov{n} \cdot \ov{n}=-1$ and $n \cdot p= \ov{n} \cdot
\ov{p}=0$, and is introduced by replacing the usual projection operators by
\beq
u(p) \overline{u}(p) & \ra & ( \not p +m) \ \frac{1+\gamma_5 \not n }{2} \ \ \
\mbox{for particles} \non \\
v(\ov{p}) \overline{v}(\ov{p}) & \ra & (\ov{\not p}-\ov{m}) \ \frac{1+\gamma_5
\ov{\not n}} {2} \ \ \ \mbox{for antiparticles}
\eeq
\nn In the production process, the polarization vector $\cp$ is defined as
\beq
{\rm d}\sigma^{\rm pol} & = & \frac{1}{2} \ {\rm d}\sigma^{\rm unpol} \ (1 +
n \cdot \cp)
\eeq

\nn Choosing the $(x,z)$ plane as the scattering plane with the electron along
the $+z$ direction, the covariant spin vector $n$ can be decomposed, in
general,
along three directions: the $F$ direction $\vec{p}$ ($\vec{n_{||}}$), the
transverse direction with respect to $\vec{p}$ but within the scattering plane
($\vec{n_\perp})$ and the transverse direction with respect to $\vec{p}$
but along a normal to the scattering plane $(\vec{n_N})$. The projection of
the spin vector along these three directions defines the corresponding degrees
of polarization
\beq
n^\mu = P_{||} n_{||}^\mu + P_\perp n_\perp^\mu + P_N n_N^\mu
\eeq
In the case where there is no CP violation, as we will assume for the process
(A1), and since the very small imaginary parts from width or loop effects can
be safely neglected, there is no polarization transverse to the scattering
plane, $P_N=0$. We can therefore set the azimuthal angle to zero and take
$p_\mu=(E,|\vec{p}|\sin \theta, 0, |\vec{p}|\cos\theta)$ where $\theta$ is the
scattering angle; in this case the components of the polarization vector are
simply
\beq
\cp_{||}=E/m~(|\vec{p}|/E,\sin\theta,0,\cos\theta) \hspace*{5mm}
& , & \hspace*{0.5cm} \cp_{\perp} =(0,\cos \theta,0,-\sin \theta)
\eeq

\nn Taking into account the polarization of both final fermions, the cross
section for the pair production, eq.~(A1), can be written as
\beq
{\rm d}{\sigma}^{\rm pol}= \frac{1}{4} \ {\rm d}{\sigma}^{\rm unp} \left(
1 + {\cal P} \cdot n + \overline{\cal P} \cdot \overline{n} + C^{\mu \nu}
n_{\mu} \overline{n}_{\nu} \right)
\eeq
with $C_{\mu \nu}$ the spin correlation. \s

\nn In the decay processes, eq.~(A2), the polarization vectors of the heavy
fermions $\cp'$, are defined similarly to eq.~(A2) but without the factor 1/2;
for instance
\beq
{\rm d} \Gamma^{\rm pol} &=& {\rm d} {\Gamma}^{\rm unpol} \
(1 + n \cdot \cp')
\eeq

\nn In the narrow width approximation, the full cross section for the
production
and subsequent decay of the heavy fermions,
\beq
\ee \ra f_0 \ov{f_0}\; f_1\; \ov{f_1} \; f_2 \; \ov{f_2}
\eeq
can be written as
\beq
{\rm d}{\sigma}= \frac{1}{4} {\rm d}\sigma^{\rm unpol} {\rm d}
\Gamma^{\rm unpol} \ {\rm d}\overline{\Gamma}^{\rm unpol}
\left( 1 + \eta_{\mu \nu}\;{\cal P}^{\mu} {\cal P'}^{\nu} +
\overline{\eta}_{\mu \nu} \overline{{\cal P}}^{\mu} \overline{{\cal P'}}^{\nu}
+ \eta_{\mu \alpha} \overline{\eta}_{\nu \beta} C^{\mu \nu}
{\cal P'}^{\alpha} \overline{{\cal P'}}^{\beta} \right)
\eeq
with
\beq
\eta_{\mu \nu}=-g_{\mu \nu} + p_\mu p_\nu / m^2 \ \ , \ \ \ \
\ov{\eta}_{\mu \nu}=-g_{\mu \nu} + \ov{p}_\mu \ov{p}_\nu / \ov{m}^2
\eeq

\nn Summing over the polarization of one fermion, e.g.~$\ov{F}$, the full cross
section  simplifies to
\beq
{\rm d}\sigma= \frac{1}{2} \ {\rm d} \sigma^{\rm unpol} \ {\rm d}
\Gamma^{\rm unpol} \left( 1+\eta_{\mu \nu} {\cal P}^{\mu}{\cal P'}^{\nu}
\right)
\eeq

\vspace*{2mm}

\nn Note that $\eta_{\mu \nu} {\cal P}^{\mu} {\cal P'}^{\nu}= \vec { {\cal P}
}^{*}.\vec {{\cal P'}^{*}}$, where the $^{*}$ refers to the components in the
rest frame of the heavy fermion. \\

\nn The above formalism, based on the factorization of the production and decay
sequences, permits an easy reconstruction of the full correlations between the
initial and the final particles from the decay of the heavy fermions, which
would be very difficult to obtain directly from the full process eq.~(A9) with
six particles in the final state. Furthermore, it is well adapted for the
setting of a Monte Carlo generator.

\end{document}